\documentclass[amssymb,nobibnotes,superscriptaddress,longbibliography,notitlepage,twocolumn]{revtex4-1}

\usepackage{enumerate,amsmath}
\usepackage[dvipdfmx]{graphicx}
\usepackage[dvipdfmx]{color}

\newcommand{\sX}{\mathsf{X}}
\newcommand{\sZ}{\mathsf{Z}}

\newcommand{\cH}{{\cal H}}
\newcommand{\bC}{\mathbb{C}}
\newcommand{\bF}{\mathbb{F}}

\newtheorem{theorem}{Theorem}

\newcommand{\Tr}{{\rm Tr}\,}

\begin{document}

\title{Verifiable fault-tolerance in measurement-based quantum computation}%
\author{Keisuke Fujii}
\address{Photon Science Center, Graduate School of Engineering,
The University of Tokyo, 2-11-16 Yayoi, Bunkyo-ku, Tokyo 113-8656, Japan}
\address{JST, PRESTO, 4-1-8 Honcho, Kawaguchi, Saitama, 332-0012, Japan}

\author{Masahito Hayashi}
\address{Graduate School of Mathematics, Nagoya University, Nagoya, 464-8602, Japan}
\address{Centre for Quantum Technologies, National University of Singapore, 117543, Singapore}

\date{\today}
\begin{abstract}
Quantum systems, in general, output data that cannot be simulated efficiently by a classical computer, 
and hence is useful for solving certain mathematical problems
and simulating quantum many-body systems. 
This also implies, unfortunately, that verification of the output of the quantum systems is not so trivial, since predicting the output is exponentially hard.
As another problem, 
quantum system is very delicate for noise and thus needs error correction.
Here we propose a framework for verification of the output of 
fault-tolerant quantum computation in the measurement-based model.
Contrast to existing analyses on fault-tolerance, 
we do not assume any noise model on the resource state,
but an arbitrary resource state is tested by using only single-qubit 
measurements to verify
whether the output of measurement-based quantum computation on it is correct or not.
The overhead for verification including classical processing
is linear in the size of quantum computation.
Since full characterization of quantum noise is exponentially hard 
for large-scale quantum computing systems,
our framework provides an efficient way of practical verification of 
experimental quantum error correction.
Moreover, the proposed verification scheme is also 
compatible to measurement-only blind quantum computation,
where a client can accept the delegated quantum computation 
even when a quantum sever makes deviation, as long as the output is correct.
\end{abstract}

\maketitle
\noindent{\it Introduction.---}
Quantum computation provides 
a new paradigm of information processing
offering both fast and secure information processing,
which could not be realized in classical computation~\cite{NC}.
Recently, a lot of experimental efforts have been paid to realize quantum computation~\cite{LJLNMO,Gibney,BK}.
There, fault-tolerant quantum computation 
with quantum error correction~\cite{NC,KF}
is inevitable to obtain quantum advantage using noisy quantum devices.

Due to the recent rapid progresses 
on experimental quantum error correction techniques~\cite{Martinis14,Martinis15,IBM15,IBM16},
there is an increasing demand on an efficient way of a performance analysis of fault-tolerant quantum computation.
In particular, in the majority of 
performance analyses of fault-tolerant quantum computation,
a specific noise model, such as {independent and identical Pauli error operation and 
some specific correlation models}, 
is assumed apriori
~\cite{Steane,Knill05,Raussendorf06,Raussendorf07b,Raussendorf07a,AF09,FY10,FY10b,AF11}.
However, in an actual experiments more general noise occurs 
{including general trace preserving completely positive (TP-CP) maps with various correlation between qubits}.
{Further, to guarantee the correctness of the output of quantum computation,
we need to care about all cases including unexpected types of errors., i.e.,we should not assume any specific error model. 
Also, in our scenario, the full tomographic approach would not work efficiently
for the increasingly many qubits.}
Unfortunately, existing fault-tolerant quantum computations
have not equipped an efficient verification scheme yet.

The aim of this paper is to develop a fault-tolerant quantum computation 
equipping a verification scheme without assuming the underlying noise model.
As properties of verifiable fault-tolerance,
we require following two conditions.
One is {\it detectability}
which means that if
the error of a quantum computer is not correctable,
such a faulty output of the quantum computation is detected 
with high probability.
The other is {\it acceptability}
which means that an appropriately constructed quantum computer
can pass the verification with high probability.
In other words,
under a realistic noise model, 
the test accepts the quantum computation 
with high probability. 
Both properties are important to characterize
performance of test in statistical hypothesis testing~\cite{textbook}.

In this paper, we develop 
verifiable fault-tolerance in measurement-based quantum computation (MBQC)
~\cite{RB01,RBB03},
which satisfies both detectability and acceptability.
We take a rather different approach to fault-tolerance
than conventional one.
We do not assume any noise model underlying,
but define a correctable set of errors 
on a resource state of MBQC
and test whether the error on a given resource state 
belongs to such a set or not.
To this end,
we employ the stabilizer test proposed in Ref.~\cite{HM},
where 
an efficient verification of MBQC
can be carried out by testing the graph state.
However, this method is not fault-tolerant lacking acceptability;
any small amount of noise on the graph state
causes rejection regardless whether or not it is correctable.
Although the paper \cite{HH} extended the stabilizer test 
to the self-testing for the measurement basis, it still has the same problem.
Therefore, we crucially extend the stabilizer test~\cite{HM} for a noisy situation,
so that we can decide whether the given resource states
belong to a set of fault-tolerant resource states or not.
Under the condition of a successful pass of the test,
the accuracy of fault-tolerant MBQC is guaranteed to be arbitrarily high
(i.e., contraposition of detectability).
The total resource required for the verification is 
linear to the size of quantum computation.
As a concrete example, 
we explicitly define a set of correctable errors on the resource state 
for topologically protected MBQC~\cite{Raussendorf06,Raussendorf07a,KF},
where we can show acceptability 
by calculating the acceptance probability concretely 
under a realistic noise model.

Note that contrast to detectability,
{the requirement of acceptability 
is unique for the verification of fault-tolerant quantum computation.
Indeed, when we can expect no error like 
the previous case \cite{HM},
we do not need fault-tolerance.
So, we could correctly judge the no error case with probability $1$,
i.e., acceptability of the test is trivially satisfied
because the stabilizer test would be passed in the no error case.
On the other hand, in verification of fault-tolerant quantum computation}
consisting of many elementary parts, each of which cannot be 
checked directly,
we have to judge whether the output of the computation is correct or not 
carefully {under an expected error model}, which
imposes the second requirement, acceptability.
{That is, to discuss acceptability, we firstly fix 
our verification method and an expected error model. Then, we calculate the acceptance probability, which corresponds to power of test 
in statistical hypothesis testing \cite{textbook}.} 


We also discuss an application of 
verifiable fault-tolerance to verification of 
blind quantum computation~\cite{BFK09,FK12,MABQC,MF12,MF13L,BKBFZW12,TFMI16}
under a quantum server's deviation
or quantum channel noise.

\noindent{\it A general setup for fault-tolerant MBQC.---}
Let us consider a generic scenario of
fault-tolerant MBQC on 
a two-colorable graph state 
composed of the black system $\cH_B$ and the white system $\cH_W$,
which are 
consist of $n_B$ and $n_W$ qubits, respectively.
Then, we have two kinds of operators
$\sX^z := X^{z_1} \otimes \cdots \otimes X^{z_{n}}$, 
$\sZ^x := Z^{x_1} \otimes \cdots \otimes Z^{x_{n}}$,
on $\cH_B \otimes \cH_W=(\bC^2)^{\otimes n}$, where $n:=n_B+n_W$.
When we restrict them to the black system $\cH_B$ (the white system $\cH_W$),
we denote $\sX^z$ and $\sZ^x$ by $\sX_B^z$ and $\sZ_B^x$ ($\sX_W^z$ and $\sZ_W^x$).
By using 
the binary-valued adjacency matrix $A$ (i.e., $(i,j)$ element is $1$ iff vertices $i$ and $j$ are connected) 
corresponding to the graph, 
the graph state $|G\rangle$ is characterized as
\begin{align}
\sX_B^{z_B} \otimes \sZ_W^{A z_B}
|G\rangle
=|G\rangle, \quad
\sX_W^{z_W} \otimes \sZ_B^{A^T z_W}
|G\rangle
=|G\rangle \label{H2}
\end{align}
for $z_B \in \bF_2^{n_B}$ and $z_W \in \bF_2^{n_W}$.
This relation explains that any error on the $Z$-basis can be converted to an error on the $X$-basis.
Then, the total space $\cH_B \otimes \cH_W$
is spanned by $\{ \sZ^{x}|G\rangle\}_{x \in \bF_2^{n}}$.
Suppose we execute a fault-tolerant MBQC quantum computation 
on the two-colorable graph state.
Then a set of correctable errors 
on the two-colorable graph state is defined such that
an ideal state $|G\rangle$ and erroneous one $\sZ^{x}|G\rangle$
result in the same computational outcome under error correction.
Such a set of errors can 
is specified as a subset $S$ of $\bF_2^{n}
=\bF_2^{n_B}\times \bF_2^{n_W}$.
The projection to the subspace is written by $\Pi_S$.
We assume that the subset $S$ is written as 
$S_B\times S_W$ by using two subsets $S_B \subset\bF_2^{n_B}$ and $S_W \subset \bF_2^{n_W}$.

\noindent{\it Test for verification of fault-tolerance.---}
\begin{figure}
\centering
\includegraphics[width=75mm]{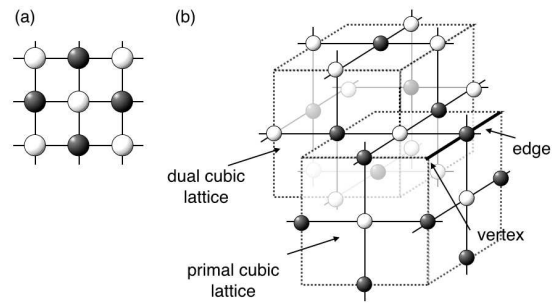}
\caption{(a) A two-colorable graph state. (b) The three dimensional two-colorable 
graph state for topologically protected measurement-based quantum computation.
}
\label{fig1}
\end{figure}
Similar to Ref.~\cite{HM}, 
we employ the following sampling protocol to verify whether the error is correctable.
Our protocol runs as follows:
\begin{itemize}
\item[1.]
Honest Bob generates $|G\rangle^{\otimes 2k+1}$, where $|G\rangle$ is
an $n$-qubit graph state on a bipartite graph $G$,
whose vertices are divided into two disjoint sets $W$ 
and $B$. (See Fig.~\ref{fig1}(a).)
Bob sends each qubit of it one by one to Alice.
Evil Bob can generate any $n(2k+1)$-qubit state $\rho$
instead of $|G\rangle^{\otimes 2k+1}$.

\item[2.]
Alice divides $2k+1$ blocks of $n$ qubits into three groups by random choice.
The first group consists of $k$ blocks of $n$ qubits each.
The second group consists of $k$ blocks of $n$ qubits each.
The third group consists of a single block of $n$ qubits.

\item[3.]
Alice uses the third group for her computation.
Other blocks are used for the test, which will be explained later.
\item[4.]
If Alice passes the test, she accepts the result of the computation performed
on the third group.
\end{itemize}

For each block of the first and second groups,
Alice performs the following test:
\begin{itemize}
\item[$T_B$]
For each block of the first group,
Alice measures qubits of $W$ in the $Z$ basis
and qubits of $B$ in the $X$ basis.
Then, she obtain $Z_W$ and $X_B$.
If $X_B+A^T Z_W \in S_B$, 
then the test is passed.

\item[$T_W$]
For each block of the second group,
Alice measures qubits of $B$ in the $Z$ basis
and qubits of $W$ in the $X$ basis.
Then, she obtain $Z_B$ and $X_W$.
If $X_W+A Z_B \in S_W$, 
then the test is passed.
\end{itemize}

\noindent
{\it Detectability and acceptability.---}
To show detectability, 
{taking account into unexpected errors,}
we obtain
the following theorem in the same way as \cite{HM}:
\begin{theorem}
\label{L1}
Assume that $\alpha > \frac{1}{2k+1}$.
If the test is passed, 
with significance level $\alpha$,
we can guarantee that 
the resultant state $\sigma$ of the third group satisfies
\begin{eqnarray}
\Tr \sigma \Pi_S \ge 
1 -\frac{1}{\alpha(2k+1)}.\label{X1}
\end{eqnarray}
\end{theorem}
(Note that the significance level is the maximum passing probability 
{when Bob erroneously generates incorrect states 
so that the resultant state $\sigma$ does not satisfy \eqref{X1}~\cite{textbook}.
That is, $1-\alpha$ expresses the minimum probability to detect such incorrect states.})
The previous study \cite{HM} considers the case with 
$S_B=\{0\}$, $S_W=\{0\}$,
and proves this special case by discussing the two kinds of binary events 
$X_B+A^T Z_W= $ or $\neq 0$
and $X_W+A Z_B= $ or $\neq 0$.
Replacing these two events by the two kinds of events
$X_B+A^T Z_W \in $ or $\notin S_B$
and $X_W+A Z_B \in $ or $\notin S_W$ in the proof given in \cite{HM},
we can show Theorem 1 with the current general form.

From the theorem and the relation between the fidelity and trace norm \cite[(6.106)]{HIKKO},
we can conclude the verifiability:
If Alice passes the test, 
she can guarantee that 
\begin{eqnarray*}
\Big|\Tr (C_x\sigma)-\Tr(C_x
\frac{\Pi_S \sigma \Pi_S}{\Tr \sigma \Pi_S}
)\Big|\le
\frac{1}{\sqrt{\alpha(2k+1)}}
\end{eqnarray*}
for any POVM $\{C_x\}$
with the significance level $\alpha$.
That is, the property of fault-tolerant quantum computation
guarantees that
the probability that the obtained computation outcome 
is different from the true computation outcome is less than
$\frac{1}{\sqrt{\alpha(2k+1)}}$.
If we take $\alpha=\frac{1}{\sqrt{2k+1}}$, for example,
this error probability is
$\frac{1}{(2k+1)^{1/4}}\to0$ if $k\to \infty$,
and therefore the verifiability is satisfied.
Note that the lower bound, $\alpha>\frac{1}{2k+1}$, of the significance level $\alpha$ is tight, since
if Bob generates $2k$ copies of the correct state $|G\rangle $ and 
a single copy of a wrong state, 
Bob can fool Alice with probability $\frac{1}{2k+1}$,
which corresponds to $\alpha=\frac{1}{2k+1}$. 
Note that the above theorem on detectability 
holds without any assumption on the underlying noise.
Note that noise in the measurements can also be taken as 
noise on the resource state, if it does not depnd on
the measurement bases.
Even if it is not the case, we can add 
noise such that the amounts of noise are the same 
for all measurement bases.

Next, we consider acceptability.
To address the success probability under a realistic noise model,
{we assume a specific application of Pauli channel 
on $\cH_B \otimes \cH_W$ as an expected error model.
That is, the error given as the distribution $P$ on the set $\bF_2^{n_B+n_W}\times \bF_2^{n_B+n_W}$
of $X$-basis errors and $Z$-basis errors.}
Then, we denote the marginal distribution with respect to the pair of 
$X$-basis errors on $B$ and $Z$-basis errors on $W$
($Z$-basis errors on $B$ and $X$-basis errors on $W$)
by $P_B$ ($P_W$).
Hence, the probability that Alice passes the test $T_B$ ($T_W$)
with one round is 
$P_B(S_B)$ ($P_W(S_W)$).
Since we apply them $2k$ rounds,
the probability to be passed is 
$P_B(S_B)^k P_W(S_W)^k$.
Hence, when the probabilities 
$P_B(S_B)$ and $P_W(S_W)$ are close to $1$,
Alice can accept the correct computation result on the third group
with high probability.

\noindent{\it Verifiable fault-tolerance for topologically protected MBQC.---} 
To show acceptability, 
below we will explain how to define a correctable set of the errors 
on a graph state.
Then, for a concrete example, 
we will calculate the acceptance probability $P_B(S_B)^k P_W(S_W)^k$
under a realistic noise model.

{In the theory of fault-tolerant quantum computation,
it is conventional that we translate fault-tolerance in the circuit model into fault-tolerance in the measurement-based model~\cite{ND,AL,RaussendorfPhD} as follows.
In the circuit model,
we can define a set of correctable (sparse) fault paths
so that the output of quantum computation does not damaged
even if any error occurs on such a fault path~\cite{AB,AGP,QEC}.
Then, translating the correctable (sparse) fault paths
in the circuit model into the measurement-based model,
we can define a correctable set of the errors on the graph state in general.}
For example, the schemes in Refs~\cite{DHNprl,DHNpra}
and Refs~\cite{Raussendorf06,Raussendorf07a} can be 
viewed as measurement-based versions of 
circuit-based fault-tolerant schemes using 
the concatenated Steane 7-qubit code~\cite{SteanNature,SteanPRA,KF,DKLP}
and the surface code with the concatenated Reed-Muller 15-qubit code~\cite{Raussendorf07b,FSG}, respectively.

\begin{figure}
\centering
\includegraphics[width=65mm]{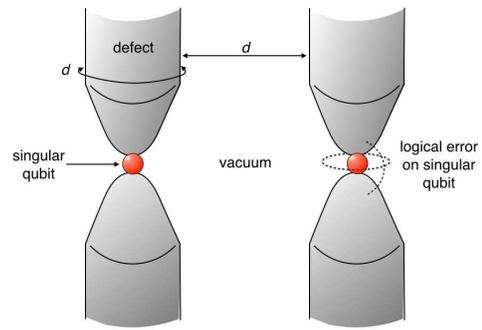}
\caption{The tubes indicate defect regions, 
in which the qubits are measured in the $Z$-basis.
Singular qubits are located in-between two defect regions,
which are measured in the $(X+Y)/\sqrt{2}$-basis 
for a transversal logical $(X+Y)/\sqrt{2}$-basis measurement. 
Other regions are vacuum, where qubits are measured in the $X$-basis
to obtain the error syndrome.
}
\label{fig2}
\end{figure}
Let us see a concrete example by using topologically 
protected MBQC~\cite{Raussendorf06,Raussendorf07a,KF},
which has been employed as a standard framework 
for fault-tolerant MBQC recently~\cite{Devitt,FT,LB,Monroe,Nemoto}.
Here we focus on the original scheme proposed in Ref.~\cite{Raussendorf06},
where the surface code and the concatenated Reed-Muller code 
are employed to perform 
two-qubit Clifford gate and single-qubit non-Pauli-basis measurements,
respectively.
In the following we will briefly sketch how 
the correctable set are defined.
A detailed description is shown in Appendix~\ref{app1}.

In the following,
we characterize the correctable sets of errors $S_B$ and $S_W$.
The errors specified by the set $S_B$,
which correspond to $X$ basis (the Pauli-$Z$ operator) on black qubits and
$Z$ basis (the Pauli-$X$ operator) on white qubits,
are detected on the priaml cubic lattice
consisting of the edges on which the black qubits
are located as shown in Fig.~\ref{fig1}(b).
Then, the error configuraion $x_B \in S_B$ 
can be associated with a set of edges on the primal cubic lattice. 
Similarly, the errors in the set $S_W$ is detected 
on the dual cubic lattice and the error configuration $x_W$ 
is associated with a set of edges on the dual cubic lattice.
In the following,
all arguments are made independently 
of black (primal lattice) and white (dual lattice) qubits.

Depending on quantum computation that Alice wants to do fault-tolerantly,
a measurement pattern is determined.
Specifically, from an analogy 
of topological quantum computation~\cite{TQC},
the sets of qubits measured in 
$X$, $Z$, and $(X+Y)/\sqrt{2}$-bases~\cite{Raussendorf06}
are called defect, vacuum, and singular qubits, respectively.
As shown in Fig.~\ref{fig2},
defect qubits shape tubes, which represent
logical degrees of freedom;
at each time slice they corresponds to the surface code 
with defects.
By braiding the defects, 
a Clifford two-qubit gate can be performed.
For the surface code,
the minimum distance decoding can be 
done by finding a shortest path connecting 
the boundary of the error chain on the cubic lattice.
Then, if the minimum distance decoding 
results in a logical operator of a weight (distance) larger than 
the code distance $d$ by
wrapping around a defect or connecting two different defects,
such an error is uncorrectable (see Appendix~\ref{app1} for the detail).
Accordingly, we can define $S^{\rm sf}_{C}$ for $C=B,W$
as the complement of them.
The code distance $d$ is chosen to be ${\rm polylog}(n')$
with $n'$ being the size of the quantum computation that 
Alice wants to do fault-tolerantly. Therefore, 
the number of qubits of the graph state 
is given by $n= {\rm poly}(n')$.

Around the singular qubits,
we still have a logical error of a weight lower than $d$
as shown in Fig.~\ref{fig2}.
Such a logical error is corrected by using another 
code, the concatenated 
Reed-Muller code.
To this end, the fault-tolerant Clifford gates
using the surface code
are further employed to encode the logical qubits
into concatenated Reed-Muller codes,
on which we can implement all Pauli-bases and $(X+Y)$-basis
measurements transversally.
The corresponding physical $(X+Y)$-basis measurements,
i.e., measurements on the singular qubits,
are depicted by red circles in Fig.~\ref{fig2}.
Then we can define the correctable set $S^{\rm rm}_{C}$ 
 of the errors for the 
concatenated Reed-Muller code recursively for $C=B,W$
as done in Ref.~\cite{AB} (see Appendix~\ref{app2} for the detail).

Since we employ two types of error correction codes as seen above,
the correctable set of the errors are defined as 
an intersection of the correctable sets $S^{\rm sf}_{C}$ 
and $S^{\rm rm}_{C}$ for the surface code 
and the concatenated Reed-Muller code, respectively, for both colors $C=B,W$.
Since both minimum distance decoding for the surface code
and recursive decoding for the concatenated code can be done 
efficiently, we can efficiently decide whether a given error pattern
$X_B +A^{T}Z_W $ ( $X_W +A^{T}Z_B $ ) are in $S_B$ ($S_W$) or not.

{\noindent{\it Acceptance probability under a typical error model.---} 
To calculate the acceptance probability,
we assume, for simplicity, the errors $\sZ^x$ ($x\in \bF^{n} $)
are distributed independently and identically for each qubit with 
probability $p$.}
It is straightforward to generalize the following argument to 
any local CPTP noise as long as the noise strength measured by the diamond norm
is sufficiently smaller than a certain threshold value~\cite{FT16}.
Then the standard counting argument
of the self-avoiding walk for the surface code~\cite{DKLP}
tells us that 
\begin{eqnarray}
P(S_{C}^{\rm sf})
> 1-{\rm poly}(n) (10p^{1/2})^d
\end{eqnarray}
for $C=B,W$.
Apparently, if $p$ is sufficiently smaller than a 
certain constant value, $P(S_{C}^{\rm sf})$ 
converges to 1 for $C=B,W$.
By considering a recursive decoding of the concatenated code,
we obtain
\begin{eqnarray}
P(S_{C}^{\rm rm})
> [1- (105^2 p_0 ^{\rm fault} )^{2^l}/105^2]^m,
\end{eqnarray}
for $C=B,W$ where $p_0 ^{\rm fault}$ 
is a logical error probability of a weight lower than $d$,
which occurs around the singular qubits.
Such a logical probability is also calculated
as a function of the physical error probaility $p$ 
by counting the number of self-avoiding walk~\cite{DKLP}
as show in Appendix~\ref{app3} Eq.~(\ref{SAW}).
The integer $m={\rm poly}(n')$ 
and $l={\rm poly}\log d)$ are the numbers of the logical $(X+Y)$-basis measurements
and the number of concatenation, respectively.
Again by using counting the number of self-avoiding walks~\cite{DKLP,FT16}
we can evaluate $p_0 ^{\rm fault}$.
By choosing $p$ smaller than a certain constant value,
$p_0 ^{\rm fault}$ becomes sufficiently small so that 
$P(S_{C}^{\rm rm})$ converges to 1 for $C=B,W$.
Since 
\begin{eqnarray}
P(S_{C}) &=& P(S_{C}^{\rm sf} \cap S_{C}^{\rm rm})
\\
&>&P( S_{C}^{\rm sf}) + P(S_{C}^{\rm rm}) -1
\end{eqnarray}
for $C=B,W$,
{the probability} $P(S_{C})$
also converges to 1 exponentially in the large $d$ limit, 
if the physical error probability $p$
is smaller than a certain constant threshold value
(see Appendix~\ref{app3} for the detailed calculation).
Since $d$ can be chosen independently of $k$,
the acceptance probability $P_B(S_B)^k P_W(S_W)^k$ converges to 1.

\noindent{\it Verifiable blind quantum computation.---}
A promising application of the proposed framework is 
verification of measurement-only blind quantum computation~\cite{MABQC}.
Suppose a quantum server generates two-colorable graph states
and sends them to a client who execute universal 
quantum computation by only single-qubit measurements,
where client employ the proposed verification.
First, our protocol is a one-way quantum communication from Bob to Alice,
and therefore, the blindness is guaranteed by the no-signaling principle
as in the protocol of Ref.~\cite{MABQC},
which contrasts to verifiable blind quantum computation~\cite{FK12,TFMI16}
of BFK (Broadbent-Fitzsimons-Kashefi) type~\cite{BFK09}.
According to Theorem~\ref{L1} (detectability) under the condition of 
acceptance the accuracy of the output is guaranteed.
Contrast to the earlier verifiable blind quantum computation~\cite{FK12,HM},
by virtue of acceptability,
the proposed verification scheme  
can accept the delegated quantum computation
even under quantum server's deviation or quantum channel noise
as long as they are correctable.
In this way, we can verify the quantum server is honest enough 
to obtain a correct output by only using single-qubit measurements.
It would be interesting to apply the proposed framework 
to quantum interactive proof systems~\cite{ABE,MFN}.

\section*{Acknowledgements}
K.F. is supported by KAKENHI No.16H02211, PRESTO, JST, CREST, JST and ERATO, JST.
MH is partially supported by 
Fund for the Promotion of Joint International Research (Fostering Joint International Research) No. 15KK0007. 
The Centre for Quantum Technologies is funded by the Singapore Ministry of Education and the National Research Foundation as part of the Research Centres of Excellence programme.
He is also grateful to Dr. Michal Hajdusek for helpful comments.

\appendix

\section{Test for topological protection}
\label{app1}
The error detection on the black vacuum qubits (edges of the primal cubic lattice) 
is executed as follows.
If there is no error on the graph state,
the outcome $m_b$ of the $X$-basis measurements 
satisfies the condition:
\begin{eqnarray}
s_{v} \equiv \bigoplus _{b \in \delta v} m_{b} =0
\end{eqnarray}
where $\bigoplus _{b \in \delta v}$ indicates 
an addition modulo two over all black qubits 
adjacent to the vertex $v$.
Depending on a given error $\sZ ^{x_B}$ ($x_B \in \bF_2^{n_B}$) on the graph state,
we can obtain the error syndrome 
$\{ s_{v}(x_B) \}$ at the vertices belonging to the defect region.
From the error syndrome,
the most likely location of the errors is estimated.
Here we employ the minimum distance decoding,
which can be done by finding a minimum path 
connecting pairs of vetices of $s_v(x_B)=1$
with the minimum-weight-perfect-matching (MWPM) algorithm~\cite{DKLP}.
Let 
\begin{eqnarray}
\bar x_B(\{s_{v}(x_B)\}) \equiv \arg \min _{x |\{s_v(x)=s_v(x_B)\} } |x| 
\end{eqnarray} 
be the estimated error location,
where $|x|$ indicates the number of 1s in a bit string $x$.
If a chain of edges specified by $x_B +\bar x_B$ 
have a nontrivial cycle in the sense of the relative homology~\cite{Raussendorf06,Raussendorf07a,Raussendorf07b},
the error correction fails.
At the defect region far from the singular qubits,
a nontrivial cycle have at least length $d$,
which is the characteristic length of the defect
determined from the required accuracy of quantum computation.
Let $n'$ be the size of the quantum computation 
that Alice wants to do fault-tolerantly.
To guarantee the accuracy of the output,
it is enough to choose the distance $d = {\rm polylog}(n')$.
Therefore, the number of the qubits of the graph state is 
$n = {\rm poly}(n')$.

Now we can define the correctable set of errors as follow:
an error location $x_B$ belongs to the correctable set $S^{\rm sf}_B \subset \bF ^{n_B}$ of the errors 
iff there exists a connected component of length $d$
in the chain of edges specified by $x_B +\bar x_B$. 
The error detection and definition of the correctable error set
$S^{\rm sf}_W$ on the white vacuum qubits 
are done in the same way but on the dual lattice.

From the test $T_B$, we know the error location $x_B$. 
Since the MWPM algorithm works in polynomial time 
in the number of vertices with $s_v=1$,
we can decide whether or not $x_B$ belongs 
to the correctable error set $S^{\rm sf}_B$.
The same argument also holds for 
the error location $x_W$ on the white vacuum qubits 
tested by $T_W$.
Therefore, we can efficiently check whether or not the errors on a
given resource belong to $ S^{\rm sf}_B \times S^{\rm sf}_W$.

\section{Test for the logical $(X+Y)/\sqrt{2}$-basis measurement}
\label{app2}
We here, for simplicity, do not employ magic state distillaion 
~\cite{Raussendorf07a,Raussendorf07b}
but encodes each logical qubit into the Reed-Muller
15-qubit code. 
Then we perform a fault-tolerant logical $(X+Y)/\sqrt{2}$-basis
measurement by transversal physical $(X+Y)/\sqrt{2}$-basis
measurements on the singular qubits
as done in Ref~\cite{Raussendorf06}.
Thereby, 
Alice can fix her strategy of 
quantum computation,
which makes easy to define the correctable set of errors for 
the test. 
Let $l$ and $m={\rm poly}(n')$ be the number of concatenation levels and 
the number of the logical $(X+Y)$-basis measurements, 
respectively.
Then we need $15^{l} m$ physical $(X+Y)/\sqrt{2}$-basis measurements,
on the singular qubits.
Note that $l=O({\rm poly}\log \log n')$ is enough to 
reduce the logical error sufficiently.
In the following, 
we the error on the graph state is 
specified by $x \in \bF_2^{n}$
by converting it into $Z$ operators 
on the graph state, $\sZ^{x} |G\rangle$.

The logical $(X+Y)/\sqrt{2}$-basis measurement
is done by physical transversal $(X+Y)/\sqrt{2}$-basis 
measurements by encoding each qubit into 
a concatenated Reed-Muller 15-qubit codes~\cite{Raussendorf06}.
This is also the case for all Pauli-basis measurements.
In the vacuum region near the singular qubits,
we have a logical error of length smaller than $d$
as shown in Fig.~\ref{fig2},
since they are not topologically protected.
Correctable error for the 
fault-tolerant logical $(X+Y)/\sqrt{2}$-basis measurement
is defined for a given error $(x_B,x_W) \in S^{\rm rm}_B \times S^{\rm rm}_W $ recursively as follows:
At physical level, which we call level-$0$,
if $x_B+\bar x_B$ or $x_W+\bar x_W$ becomes a logical error for a
singular qubit,
the level-$0$ (singular) qubit is labeled to be faulty.
At $l'$th concatenation level,
if the level-$l'$ logical qubit consisting of 15 level-$(l'-1)$ logical qubits
encoded in the Reed-Muller 15-qubit code
has two or more faulty level-$(l'-1)$ logical qubits,
the level-$l'$ logical qubit is labeled to be faulty.
At the highest level $l'=l$,
if no level-$l$ logical qubit is faulty,
the given error $(x_B,x_W)$ belongs to the correctable set 
$S_B^{\rm rm} \times S^{\rm rm}_W $.

\section{Acceptance probability}
\label{app3}
\begin{figure}
\centering
\includegraphics[width=50mm]{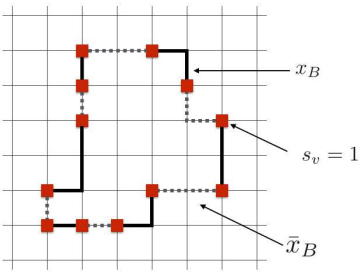}
\caption{Actual error $x_B$ and estimated one $\bar x_B$ are denoted by solid and dotted lines, respectively.
The vertices (error syndrome) of $s_v=1$ are denoted by red squares. The 3D
 lattice(spatial two dimensions and time-like one dimension) is depicted as if it is two dimensional (one dimension for both spatial and time-like axe). }
\label{fig3}
\end{figure}
Let us first consider the pass probability 
of the test for topological protection.
The error $x_B$ is rejected if $x_B+\bar x_B$
contains a connected component of length at least $d$.
Such a probability is calculated~\cite{DKLP} to be
\begin{eqnarray}
&&\sum _{\nu=d} \sum _{\mu=\nu/2}^{\nu}
n(6/5)\cdot 5^{\nu} 
\left(\begin{array}{c}
\nu
\\
\mu
\end{array}\right)
p^{\mu} (1-p)^{n-\mu}
\\
&<&
\sum _{\nu=d} 
n(6/5)\cdot (10p^{1/2})^\mu
\\
&<& {\rm poly}(n) (10p^{1/2})^d.
\end{eqnarray}
Therefore, if $p$ is sufficiently smaller than
a constant value, the rejection probability 
is exponentially suppressed.

Next we consider the test for the 
logical $(X+Y)/\sqrt{2}$-basis measurement.
Let $p_{0}^{\rm fault}$ be the 
probability that a level-0 (singular) qubit is faulty.
$p_{0}^{\rm fail}$ is evaluated 
in a similar way to the previous case 
for the topological protection but
we have to count logical errors 
consist of the chains of length lower than $d$:
\begin{eqnarray}
p_{0}^{\rm fault} =  \sum _{\nu=1}^{d}  \sum _{\mu=\nu/2}^{\nu}
C_{\nu}
\left(\begin{array}{c}
\nu
\\
\mu
\end{array}\right)
p^{\mu} (1-p)^{n-\mu},
\label{SAW}
\end{eqnarray}
where $C_\nu$ is the number of 
chains of length $\nu$ that 
contribute to the logical error of 
length $\nu$.
$C_\nu$ is counted in Ref.~\cite{FT16} rigorously
up to $\nu=14$,
which indicates that we can reduce $p_{0}^{\rm fault}$
by decreasing $p$ sufficiently.

The probability $p_{l'}^{\rm fault}$ 
of obtaining the level-$l'$ faulty qubit
is given recursively by
\begin{eqnarray}
p_{l'}^{\rm fault} &<& \sum _{r=2}^{15} ( p_{l'-1}^{\rm fault} )^r (1-p_{l'-1}^{\rm fault} )^{15-r}
\\
&=& (7\cdot 15)^2 (p_{l'-1}^{\rm fault})^2.
\end{eqnarray}
The we obtain
\begin{eqnarray}
p_{l}^{\rm fault}=  (105^2 p_0 ^{\rm fault} )^{2^l}/105^2.
\end{eqnarray}
The probability to obtain
no faulty level-$l$ logical qubit 
at the highest level 
is given by $(1-p_{l}^{\rm fault})^{m}$.

Accordingly, 
if $p_{0}^{\rm fault}$ is sufficiently smaller than
$1/(7\cdot 15)^2$, we can reduce the rejection probability 
of the test for the fault-tolerant 
logical $(X+Y)\sqrt{2}$-basis measurement.

Since $m={\rm poly}(n')$,
it is sufficient to chose $d= {\rm poly} (\log n')$
and $l = {\rm poly} (\log d)$,
which are independent of $2k+1$,
the number of the samples of the graph state.
Therefore, 
in the large $d$ limit for a given $n'$,
we can reduce the logical error probability polynomially,
and hence amplify the acceptance probability 
$P_B(S_{B})^k P_W(S_{w})^k$ arbitrarily close to 1.

\end{document}